\documentclass[a4paper,11pt]{article}
\usepackage{pos}
\usepackage{float}
\usepackage[symbol]{footmisc}
\renewcommand{\thefootnote}{\fnsymbol{footnote}}
\newcommand{\KK}{${\cal KK}$}
\def\rQCED{{\rm QCED}}
\newcommand{\qb}{{\bar{q}}}
\newcommand{\sfac}{\mathfrak{s}}
\title{New Results from IR-Improved Amplitude-Based Resummation in Quantum Field Theory}

\author*[a]{B.F.L. Ward}
\author[b]{S. Jadach\footnote[2]{Deceased.}}
\author[c]{W. Placzek}
\author[b]{M. Skrzypek}
\author[b]{Z.A. Was}
\author[d]{S.A. Yost}

\affiliation[a]{Department of Physics, Baylor University,\\
 One Bear Place \# 97316, Waco, TX 76798-7316, USA}

\affiliation[b]{Institute of Nuclear Physics, Polish Academy of Sciences,\\
ul. Radzikowskiego 152, 31-342 Krakow, Poland}

\affiliation[c]{Institute of Applied Computer Science, Jagiellonian University,\\
ul. Prof. Stanisława Lojasiewicza 11, 30-348 Krakow, Poland}

\affiliation[d]{Department of Physics, The Citadel,\\
171 Moultrie Street, Charleston, SC 29409, USA}

\emailAdd{bfl\_ward@baylor.edu}

\abstract{There is a continuing effort to support and prepare the precision physics programs for the present and planned future colliders such as HL-LHC, FCC, CLIC, CEPC, and CPPC. We discuss new results from IR-improved amplitude-based resummation in quantum field theory relevant to such support and preparation with some emphasis on the interplay between soft and collinear resummation algebras.}
\centerline{BU-HEPP-23-02, Oct., 2023}
\FullConference{16th International Symposium on Radiative Corrections: Applications of Quantum Field Theory to Phenomenology (
RADCOR2023)\\
28th May - 2nd June, 2023\\
Crieff, Scotland, UK\\}

\begin{document}
\maketitle

\section{In Memoriam}
\begin{figure}[h]
\begin{center}
\setlength{\unitlength}{1in}
\begin{picture}(6,4.5)(0,0)
\put(0.5,0){\includegraphics[width=5in]{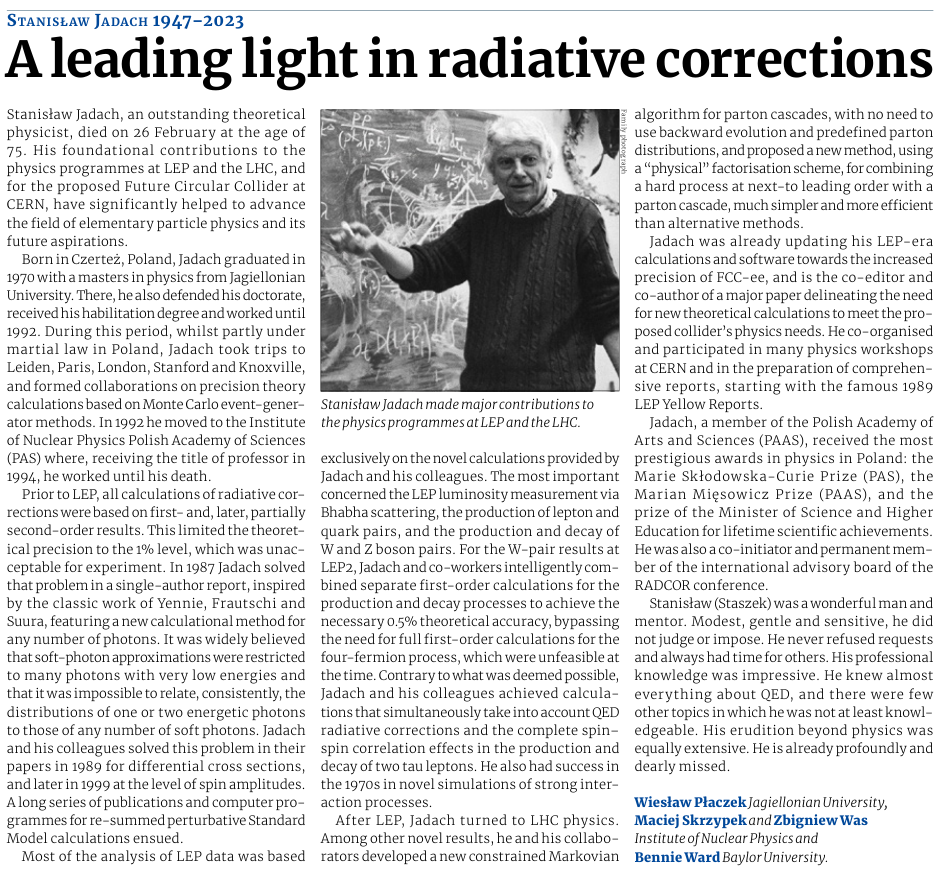}}
\end{picture}
\end{center}
\vspace{-5mm}
\caption{{\it CERN Courier} obituary for the late Prof. Stanislaw Jadach.}
\label{fig1}
\end{figure}

Sadly, my (BFLW) close friend and collaborator, Prof. Stanislaw Jadach, passed away suddenly on February 26, 2023. His {\it CERN Courier}\footnote{{\it CERN Courier}, May/June 2023 issue, p.59.} obituary is reproduced here in Fig.~\ref{fig1}. 
He was a pioneering member of the RADCOR International Advisory Board and he was the Chair of the Local Organizing Committee when the Institute of Nuclear Physics hosted the 1996 RADCOR in Krakow, Poland. His many outstanding contributions to our field helped to keep our field alive. We all miss him dearly. This contribution is dedicated in memoriam to him.\par
\section{Introduction}
The future of precision quantum field theory is dictated by the planned future colliders: FCC~\cite{fccwksp2019}, CLIC~\cite{CLIC-burrows}, ILC~\cite{ILC-behnke} CEPC~\cite{CEPC-gao}, CPPC~\cite{CPPC-tang}, $\ldots$. For example, in the case of the FCC-ee, factors of improvement from $\sim 5$ to $\sim 100$  are needed from theory. Resummation is key to such improvements in many cases. In this context, we discuss here the amplitude-based resummation following from the methodology of Yennie, Frautschi, and Suura (YFS) in Ref.~\cite{yfs:1961} 
as it has been realized Refs.~\cite{Jadach:1993yv,bhlumi4:1996,Jadach:1999vf-sh,Jadach:2000ir,Jadach:2013aha,kkmchh1-sh,Jadach:2022vf}. This approach to resummation treats the resummation of infrared (IR) singularities to all orders in the loop expansion. One of us (BFLW) has extended~\cite{bflw:hrdcrrnA} this methodology to non-Abelian gauge theories as well.\par
\begin{figure}[h]
\begin{center}
\setlength{\unitlength}{1in}
\begin{picture}(6.5,3.0)(0,0)
\put(0,0.5){\includegraphics[width=3.1in,height=2.6in]{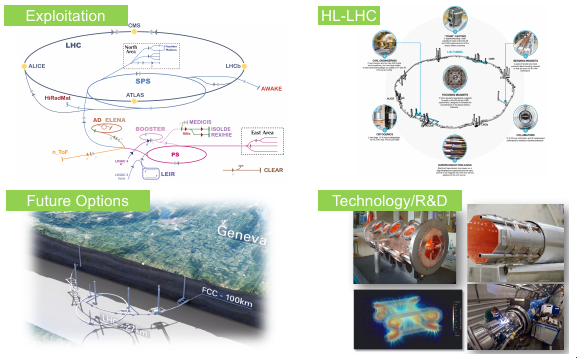}}
\put(3.0,0.5){\includegraphics[width=3.1in,height=2.6in]{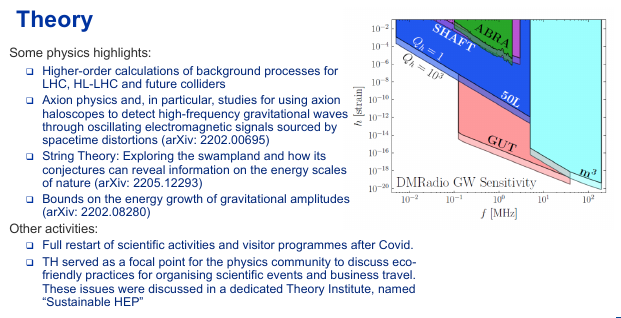}}
\put(1.5,0.25){(a)\hspace{2.8in}(b)}
\end{picture}
\end{center}
\vspace{-10mm}
\caption{\baselineskip=11pt Excerpts from Ref.~\cite{fabiola-1-10-23} on the state of CERN: (a), Future Options and R\& D; (b), Theory highlights.}
\label{fig2}
\end{figure}

It is encouraging that the need for precision theory for the future collider physics objectives is appreciated at world-leading laboratories such as CERN, as we illustrate in Fig.~\ref{fig2}
with excerpts from Ref.~\cite{fabiola-1-10-23}. The future options for CERN are shown featuring the FCC and the important role of higher-order calculations for its background processes
is a theory highlight. We can hope that the funding agencies take note of the implied connection.\par
The fundamental attraction of the YFS approach is that there is no limit in principle to the precision one may achieve as long as one calculates the corresponding hard radiation residuals to the desired order in the respective coupling constant. This should be contrasted with other methods of resummation in which various degrees of freedom are integrated out to allow the resummation
thereby engendering an intrinsic uncertainty in observables which require the reinstatement of those degrees of freedom; for, the reinstatement can only be done approximately.\par
More specifically, YFS methods in QED are exact to all orders in the infrared limit but treat collinear big logs perturbatively in the hard photon residuals. DGLAP-based collinear factorization as
recently done at the next-leading log level in Refs.~\cite{frixione-2019,bertone-2019,frixione-2021,bertone-2022}, for example, treats collinear logs to all orders but has a non-exact infrared limit. In what follows, after briefly reviewing in the next Section the elements of exact amplitude resummation theory which we employ, we first present (in Section 4) new results for precision collider physics and pioneering results in quantum gravity using the usual YFS methods. We then investigate in Section 5 improving the collinear limit of YFS theory. 
A key point is the following: Exact amplitude-based resummation realized on an event-by-event basis gives enhanced precision for a given level of exactness and this is essential for future precision physics as exemplified by CERN.\par 
\section{Brief Review of Exact Amplitude-Based Resummation Theory}
As it is still not generally familiar, we include here a synopsis of exact amplitude-based resummation theory. The master formula that exhibits the theory is as follows:
{\small
\begin{eqnarray}
&d\bar\sigma_{\rm res} = e^{\rm SUM_{IR}(QCED)}
   \sum_{{n,m}=0}^\infty\frac{1}{n!m!}\int\prod_{j_1=1}^n\frac{d^3k_{j_1}}{k_{j_1}} \cr
&\prod_{j_2=1}^m\frac{d^3{k'}_{j_2}}{{k'}_{j_2}}
\int\frac{d^4y}{(2\pi)^4}e^{iy\cdot(p_1+q_1-p_2-q_2-\sum k_{j_1}-\sum {k'}_{j_2})+
D_\rQCED} \cr
&{\tilde{\bar\beta}_{n,m}(k_1,\ldots,k_n;k'_1,\ldots,k'_m)}\frac{d^3p_2}{p_2^{\,0}}\frac{d^3q_2}{q_2^{\,0}},
\label{subp15b}
\end{eqnarray}}
where the {\em new}\footnote{The {\em non-Abelian} nature of QCD requires a new treatment of the corresponding part of the IR limit~\cite{Gatheral:1983} so that we usually include in ${\rm SUM_{IR}(QCED)}$ only the leading term from the QCD exponent in Ref.~\cite{Gatheral:1983} -- the remainder is included in the residuals $\tilde{\bar\beta}_{n,m}$ .}(YFS-style) residuals   
{$\tilde{\bar\beta}_{n,m}(k_1,\ldots,k_n;k'_1,\ldots,k'_m)$} have {$n$} hard gluons and {$m$} hard photons. The new residuals and the  infrared functions ${\rm SUM_{IR}(QCED)}$ and ${ D_\rQCED}$ are defined in Ref.~\cite{mcnlo-hwiri,mcnlo-hwiri1}.  As explained in Ref.~\cite{mcnlo-hwiri,mcnlo-hwiri1}, parton shower/ME matching engenders the replacements {$\tilde{\bar\beta}_{n,m}\rightarrow \hat{\tilde{\bar\beta}}_{n,m}$}, which allow us to connect with  MC@NLO~\cite{mcnlo,mcnlo1}, via the basic formula{\small
\begin{equation}
{d\sigma} =\sum_{i,j}\int dx_1dx_2{F_i(x_1)F_j(x_2)} d\hat\sigma_{\rm res}(x_1x_2s).
\label{bscfrla}
\end{equation}}
\par
New results in precision LHC and FCC physics have been obtained using Eq.(\ref{subp15b}). One of us (BFLW) has extended the latter equation to general relativity as an approach to quantum gravity.  In each respective application, our new results are accompanied with new perspectives. We discuss such new results and perspectives in the next Section.\par
\section{New Perspectives for Precision Physics: High Energy Colliders and Quantum Gravity} 
\begin{figure}[h!]
\begin{center}
\setlength{\unitlength}{1in}
\begin{picture}(6,2.0)(0,0)
\put(0.5,0){\includegraphics[width=5in]{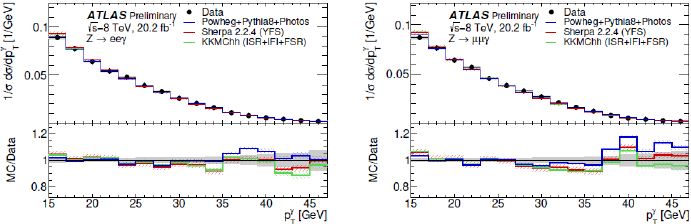}}
\end{picture}
\end{center}
\vspace{-5mm}
\caption{ATLAS analysis of $Z/\gamma$ production at $8$ TeV.}
\label{fig3}
\end{figure}

The realization of eq.(\ref{subp15b}) in the MC event generator \KK{MC}-hh~\cite{kkmchh1-sh} by four of us (SJ, BFLW, ZAW, SAY) allows a new perspective on the expectations for precision physics for the Standard Theory
EW interactions at HL-LHC. This is illustrated by the plots in Fig.~\ref{fig3} in the ATLAS analysis~\cite{atlas-conf-2022-046} of $Z\gamma$ production at 8 TeV.
The data are compared to the Powheg-Pythia8-Photos~\cite{powheg-org,powheg-orga,powheg1,powhega,Sjostrand:2007gs-sh,Golonka:2006tw}, Sherpa2.2.4(YFS)~\cite{sherpa,sherpa-2.2}, and \KK{MC}-hh predictions for the $\gamma p_T$ spectrum. At this level of the uncertainties in the data at this point, 
all three predictions are in reasonable agreement with the data. At HL-LHC, with ~10 times the current statistics, a precision test against the theories will obtain.\par
\begin{figure}[h!]
\begin{center}
\setlength{\unitlength}{1in}
\begin{picture}(6,2.0)(0,0)
\put(0.5,0){\includegraphics[width=5in]{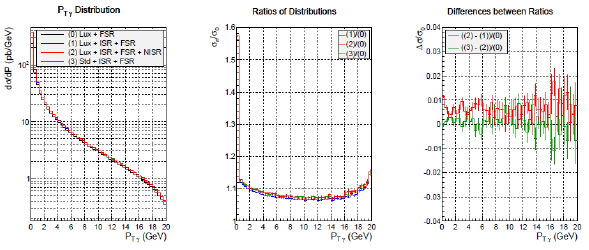}}
\end{picture}
\end{center}
\vspace{-5mm}
\caption{The distribution for $P_{T_\gamma}$ of the photon for which it is greatest for events with at least one photon and each lepton having $ p_{T\ell}> 25 GeV, \eta_\ell< 2.5$ calculated with (0) FSR only (black). (1) FSR + ISR (blue). and (2) FSR + ISR with NISR (red) for NNPDF3.1-LuxQED NLO PDFs. For comparison, (3) shows FSR + ISR with ordinary NNPDF3.1 NLO PDFs (green). The center graph shows
ISR on/off ratios (1)/(0) (blue),(2)/(0) (red) and (3)/(0) (green). The right-hand graph shows the fractional differences ((1)- (2))/(0) in red and ((2)- (3))/(0) in green.}
\label{fig4}
\end{figure}

 
An important issue is the effect of QED contamination in non-QED PDFs. In a new perspective toward this issue~\cite{jad-yost-afb,ichep2022-say,sjetaltoappear} we use Negative ISR (NISR) evolution to address the size of this contamination directly. We have, using a standard notation for PDFs and cross sections, the cross section representation
\begin{equation}
\begin{split}
\sigma(s)&=
\frac{3}{4}\pi\sigma_0(s)\!\!\!
\sum_{q=u,d,s,c,b} \int d\hat{x}\;  dz dr dt \; \int dx_q dx_\qb\; 
\delta(\hat{x}-x_qx_\qb z t)
\\&\times
f^{h_1}_q(   s\hat{x}, x_q) 
f^{h_2}_\qb( s\hat{x}, x_\qb) \;
 \rho_I^{(0)}\big(\gamma_{Iq}(s\hat{x}/m_q^2),z\big)\; 
 \rho_I^{(2)}\big(-\gamma_{Iq}(Q_0^2/m_q^2),t\big)\; 
\\&\times
\sigma^{Born}_{q\qb}(s\hat{x}z)\;
\langle W_{MC} \rangle,
\label{eq:kkhhsigmaPru}
\end{split}
\end{equation}
which includes an extra convolution with the well known second order exponentiated ISR
``radiator function'' $\rho_I^{(2)} $ with the negative evolution time argument
$-\gamma_{Iq}(Q_0^2/m_q^2)$ defined in Ref.~\cite{jad-yost-afb}. The QED below $Q_0$ is thus removed. We illustrate this in Fig.~\ref{fig4} from Ref.~\cite{ichep2022-say} 
for the $P_{T_\gamma}$ for the photon for which it is the largest in $Z\gamma^*$ production and decay to lepton pairs at the LHC at $8 ~TeV$ for
cuts as described in the figure. As we see in the figure, the results show that the effect of QED contamination in non-QED
PDFs is below the errors on the PDFs in agreement with arguments in Ref.~\cite{kkmchh2}.\par
\begin{figure}[h]
\begin{center}
\setlength{\unitlength}{1in}
\begin{picture}(6.5,3.0)(0,0)
\put(0,0.5){\includegraphics[width=3.0in,height=1.5in]{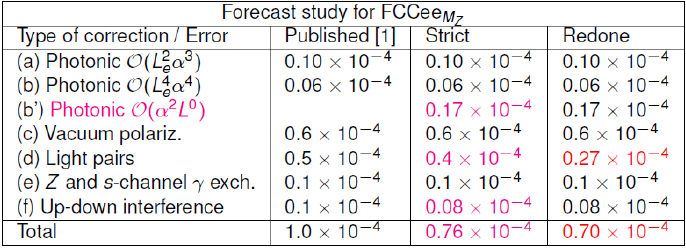}}
\put(3.05,0.5){\includegraphics[width=3.0in,height=1.5in]{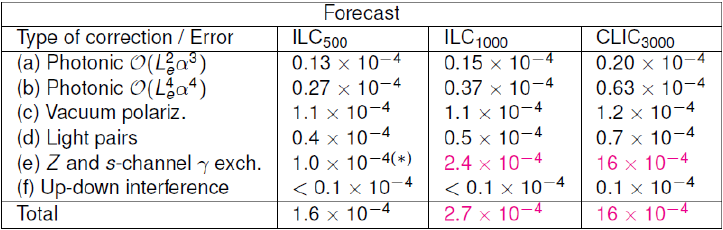}}
\put(1.5,0.25){(a)\hspace{2.8in}(b)}
\end{picture}
\end{center}
\vspace{-10mm}
\caption{\baselineskip=11pt Current purview on luminosity theory errors: (a), FCC-ee at $M_Z$; (b), proposed higher energy colliders}
\label{fig5}
\end{figure}

For the planned EW/Higgs factories, five of us (SJ, WP, MS, BFLW, SAY) have discussed in Refs.~\cite{Jadach:2018jjo,Jadach:2021ayv-sh,fcc2023wkshpms} the new perspectives for the BHLUMI~\cite{bhlumi4:1996} luminosity theory error. This new perspective is illustrated in Fig.~\ref{fig5}~\cite{fcc2023wkshpms} wherein we show the current purview for the FCC-ee at $M_Z$ and that for the proposed higher energy colliders. 
In addition to the improvements at $M_Z$ shown in Fig.~\ref{fig5}(a) to 0.007\%, the use of the results in Ref.~\cite{fjeger-fccwksp2019} together with lattice methods~\cite{latt1,latt2}
opens the possibility that item (c) in Fig~\ref{fig5}(a) could be reduced by a factor of ~ 6\footnote{The formula to be studied is $\Delta\alpha_{had}(t)=\Delta\alpha_{had}(-Q^2_0)|_{lat}+[\Delta\alpha_{had}(t)-\Delta\alpha_{had}(-Q^2_0)]|_{pQCDAdler}$ with {\it lat} denoting the methods of Refs.~\cite{latt1,latt2} and {\it pQCDAdler} denoting the methods of Ref.~\cite{fjeger-fccwksp2019}.}. \par
In Refs.~\cite{ward:2013dkunv,ijmpa2018} one of us (BFLW) has shown that amplitude-based resummation applied to quantum gravity tames its UV divergences. One of the many 
consequences is, using a standard notation,  
\begin{equation}
\begin{split}
\rho_\Lambda(t_0)&\cong \frac{-M_{Pl}^4(1+c_{2,eff}k_{tr}^2/(360\pi M_{Pl}^2))^2}{64}\sum_j\frac{(-1)^Fn_j}{\rho_j^2}\cr
          &\qquad\quad \times \frac{t_{tr}^2}{t_{eq}^2} \times (\frac{t_{eq}^{2/3}}{t_0^{2/3}})^3\cr
    &\cong \frac{-M_{Pl}^2(1.0362)^2(-9.194\times 10^{-3})}{64}\frac{(25)^2}{t_0^2}\cr
   &\cong (2.4\times 10^{-3}eV)^4.
\end{split}
\label{eq-rho-expt}
\end{equation}
Here, $t_0$ is the age of the universe and we take it to be $t_0\cong 13.7\times 10^9$ yrs and $t_{tr}\sim 25 t_{Pl}$~\cite{reuter2,ward:2013dkunv,ijmpa2018} is the transition time between the Planck regime and the classical Friedmann-Robertson-Walker(FRW) regime.
In the estimate in (\ref{eq-rho-expt}), the first factor in the second line comes from the radiation dominated period from
$t_{tr}$ to $t_{eq}$ and the second factor
comes from the matter dominated period from $t_{eq}$ to $t_0$.
The estimate in (\ref{eq-rho-expt}) is close to the experimental result~\cite{pdg2008}\footnote{See also Ref.~\cite{sola2,sola3} for analyses that suggest 
a value for $\rho_\Lambda(t_0)$ that is qualitatively similar to this experimental result.} 
$\rho_\Lambda(t_0)|_{\text{expt}}\cong ((2.37\pm 0.05)\times 10^{-3}eV)^4$. 
\par
\section{Improving the Collinear Limit in YFS Theory}
It is known~\cite{Jadach:2023cl1} that in the usual YFS theory the virtual infrared function $B$ in the s-channel resums (exponentiates) the non-infrared term
$\frac{1}{2}Q_e^2{\alpha\over\pi} L$ in $e^+(p_2)\;e^-(p_1)\rightarrow \bar{f}(p_4)\;e^-(p_3)$ using an obvious notation where the respective big log is $L = \ln(s/m_e^2)$ when $s=(p_1+p_2)^2$ is the center-of-mass energy squared. It is also known from Ref.~\cite{gribv-lptv:1972} that the term $\frac{3}{2}Q_e^2{\alpha\over\pi} L$ exponentiates -- see also Refs.~\cite{frixione-2019,bertone-2019,frixione-2021,bertone-2022} for recent developments in
attendant collinear factorization approach. Does the YFS theory allow an extension that would also exponentiate the latter term? In Ref.~\cite{Jadach:2023cl1}, three of us (SJ, BFLW, ZAW) have answered this question in the affirmative.\par
Specifically, we find that the virtual infrared function $B$ in the s-channel can be extended to
\begin{equation}
\begin{split}
B_{CL}&\equiv B+{\bf \Delta B}\\
          &= \int {d^4k\over k^2} {i\over (2\pi)^3} 
                       \bigg[\bigg( {2p-k \over 2kp-k^2} - {2q+k \over 2kq+k^2} \bigg)^2{\bf -\frac{4pk-4qk}{(2pk-k^2)(2qk+k^2)}}\bigg],
\end{split}
\end{equation}
while the real infrared function $\tilde{B}$ can be extended to 
\begin{equation}
\begin{split}
\tilde{B}_{CL} &\equiv \tilde{B}+{\bf \Delta\tilde{B}}\\
 &=\frac{-1}{8\pi^2}\int\frac{d^3k}{k_0}\bigg\{\large(\frac{p_1}{kp_1} - \frac{p_2}{kp_2}\large)^2 +{\bf \frac{1}{kp_1}\large(2 -\frac{kp_2}{p_1p_2}\large)}\\
                                          &\qquad\qquad+{\bf \frac{1}{kp_2}\large(2 -\frac{kp_1}{p_1p_2}\large)}\bigg\},
\end{split}
\label{eq-real2}
\end{equation}
where the extensions are indicated in boldface in an obvious notation. The YFS infrared algebra is unaffected by these extensions while the $B_{CL}$ does exponentiate the entire $\frac{3}{2}Q_e^2{\alpha\over\pi} L$ term and the $\tilde{B}_{CL}$ does carry the respective collinear big log of the exact result in Ref.~\cite{berends-neerver-burgers:1988} in the soft regime.\par 
The corresponding collinear extension of the CEEX soft eikonal amplitude factor defined in Ref.~\cite{ceex2:1999sh}
for the photon polarization $\sigma$ and $e^-$  helicity $\sigma'$ is given by 
\begin{equation}
\begin{split}
\sfac_{CL,\sigma}(k) = \sqrt{2}Q_ee\bigg[-\sqrt{\frac{p_1\zeta}{k\zeta}}\frac{<k\sigma|\hat{p}_1 -\sigma>}{2p_1k}
    +{\bf \delta_{\sigma'\;-\sigma}\sqrt{\frac{k\zeta}{p_1\zeta}}\frac{<k\sigma|\hat{p}_1 \sigma'>}{2p_1k}}\\
    + \sqrt{\frac{p_2\zeta}{k\zeta}}\frac{<k\sigma|\hat{p}_2 -\sigma>}{2p_2k}+{\bf \delta_{\sigma' \sigma}\sqrt{\frac{k\zeta}{p_2\zeta}}\frac{<\hat{p}_2 \sigma'|k -\sigma>}{2p_2k}}\bigg],
\end{split}
\label{eq-real4}
\end{equation}
where from Ref.~\cite{ceex2:1999sh} $\zeta\equiv (1,1,0,0)$ for our choice for the respective auxiliary vector in our Global Positioning of Spin (GPS)~\cite{gps:1998} spinor conventions with the consequent definition $\hat{p}= p - \zeta m^2/(2\zeta p)$
for any four vector $p$ with $p^2 = m^2.$ The collinear extension terms are again indicated in boldface.\par
We expect these extended infrared functions to give in general a higher precision for a given level of exactness~\cite{bflwetaltoappear}.\par
\baselineskip=14pt
\bibliography{Tauola_interface_design}{}

\providecommand{\href}[2]{#2}\begingroup\begin{thebibliography}{10}

\bibitem{fccwksp2019}
A.~Blondel {\em et al.}, {\em CERN Yellow Reports: Monographs, CERN-2020-003}
  (2020) \href{http://www.arXiv.org/abs/1905.05078}{{\tt 1905.05078}}.

\bibitem{CLIC-burrows}
{P.N. Burrows}, {\em PoS} {\bf ICHEP2020} (2020)
683.

\bibitem{ILC-behnke}
{T. Behnke} {\em et al.} (2013) \href{http://www.arXiv.org/abs/1306.6327}{{\tt
  1306.6327}}.

\bibitem{CEPC-gao}
{J. Gao} {\em et al.}, {\em PoS} {\bf ICHEP2020} (2020)
686.

\bibitem{CPPC-tang}
{J.Tang}, {\em Front. in Phys.} {\bf 10} (2022) 828878.

\bibitem{yfs:1961}
D.~R. Yennie, S.~Frautschi, and H.~Suura, {\em Ann. Phys. (NY)} {\bf 13} (1961)
  379 -- 452.

\bibitem{Jadach:1993yv}
S.~Jadach, B.~F.~L. Ward, and Z.~W\c{a}s, {\em Comput. Phys. Commun.} {\bf 79}
  (1994)
503--522.

\bibitem{bhlumi4:1996}
S.~Jadach, W.~Placzek, E.~Richter-W\c{a}s, B.~F.~L. Ward, and Z.~W\c{a}s, {\em
  Comput. Phys. Commun.} {\bf 102} (1997)
229.

\bibitem{Jadach:1999vf-sh}
S.~Jadach, B.~F.~L. Ward, and Z.~W\c{a}s, {\em Comput. Phys. Commun.} {\bf 130}
  (2000)
260--325.

\bibitem{Jadach:2000ir}
S.~Jadach, B.~F.~L. Ward, and Z.~W\c{a}s, {\em Phys. Rev.} {\bf D63} (2001)
  113009,
\href{http://www.arXiv.org/abs/hep-ph/0006359}{{\tt hep-ph/0006359}}.

\bibitem{Jadach:2013aha}
S.~Jadach, B.~F.~L. Ward, and Z.~W\c{a}s, {\em Phys. Rev.} {\bf D88} (2013),
  no.~11 114022,
\href{http://www.arXiv.org/abs/1307.4037}{{\tt 1307.4037}}.

\bibitem{kkmchh1-sh}
S.~Jadach, B.~F.~L. Ward, Z.~W\c{a}s, and S.~Yost, {\em Phys. Rev. D} {\bf 99}
  (2019) 076016.

\bibitem{Jadach:2022vf}
S.~Jadach, B.~F.~L. Ward, Z.~W\c{a}s, S.~Yost, and A.~Siodmok, {\em Comput.
  Phys. Commun.} {\bf 283} (2023) 108556,
\href{http://www.arXiv.org/abs/2204.11949}{{\tt 2204.11949}}.

\bibitem{bflw:hrdcrrnA}
{B.F.L. Ward}, {\em Mod. Phys. Lett. A} {\bf 31} (2016) 1650126.

\bibitem{fabiola-1-10-23}
F.~Gianotti {\em et al.}, {in {\it Looking forward to the New Year ...}, CERN,
  Geneva, Switzerland, Jan. 10, 2023}.

\bibitem{frixione-2019}
{F. Frixione}, {\em J. High Energy Phys.} {\bf 1911} (2019) 158,
  \href{http://www.arXiv.org/abs/1909.03886}{{\tt 1909.03886}}.

\bibitem{bertone-2019}
{V. Bertone} {\em et al.}, {\em J. High Energy Phys.} {\bf 2003} (2020) 135,
  \href{http://www.arXiv.org/abs/1911.12040}{{\tt 1911.12040}}.

\bibitem{frixione-2021}
{F. Frixione}, {\em J. High Energy Phys.} {\bf 2021} (2021) 180,
  \href{http://www.arXiv.org/abs/2105.06688}{{\tt 2105.06688}}.

\bibitem{bertone-2022}
{V. Bertone} {\em et al.}, \href{http://www.arXiv.org/abs/2207.03265}{{\tt
  2207.03265}}.

\bibitem{Gatheral:1983}
J.~Gatheral, {\em Phys. Lett.} {\bf 133B} (1983) 90.

\bibitem{mcnlo-hwiri}
{ S.K. Majhi} {\em et al.}, {\em Phys. Lett. B} {\bf 719} (2013) 367,
  \href{http://www.arXiv.org/abs/hep-ph/1208.4750}{{\tt hep-ph/1208.4750}}.

\bibitem{mcnlo-hwiri1}
{ A. Mukhopadhyay and B.F.L. Ward}, {\em Mod. Phys. Lett. A} {\bf 31} (2016)
  1650063, \href{http://www.arXiv.org/abs/hep-ph/1412.8717}{{\tt
  hep-ph/1412.8717}}.

\bibitem{mcnlo}
{S. Frixione and B.Webber}, {\em J. High Energy Phys.} {\bf 0206} (2002) 029.

\bibitem{mcnlo1}
{ S. Frixione} {\em et al.}, {\em J. High Energy Phys.} {\bf 1101} (2011) 053,
  \href{http://www.arXiv.org/abs/hep-ph/1010.0568}{{\tt hep-ph/1010.0568}}.

\bibitem{atlas-conf-2022-046}
G.~Aad {\em et al.}, ATLAS CONF note STDM-2022-046, 2022.

\bibitem{powheg-org}
P.~Nason, {\em J. High Energy Phys.} {\bf 11} (2004) 040.

\bibitem{powheg-orga}
S.~Frixione, P.~Nason, and C.~Oleari, {\em J. High Energy Phys.} {\bf 11}
  (2007) 070, \href{http://www.arXiv.org/abs/hep-ph/0709.2092}{{\tt
  hep-ph/0709.2092}}.

\bibitem{powheg1}
S.~Alioli, P.~Nason, O.~Oleari, and E.~Re, {\em J. High Energy Phys.} {\bf 06}
  (2010) 043, \href{http://www.arXiv.org/abs/hep-ph/1002.2581}{{\tt
  hep-ph/1002.2581}}.

\bibitem{powhega}
S.~Alioli, P.~Nason, O.~Oleari, and E.~Re, {\em J. High Energy Phys.} {\bf
  0807} (2008) 060, \href{http://www.arXiv.org/abs/0805.4802}{{\tt 0805.4802}}.

\bibitem{Sjostrand:2007gs-sh}
T.~Sjostrand, S.~Mrenna, and P.~Skands, {\em Comput. Phys. Commun.} {\bf 178}
  (2008)
852--867.

\bibitem{Golonka:2006tw}
P.~Golonka and Z.~W\c{a}s, {\em Eur. Phys. J.} {\bf C50} (2007) 53--62,
\href{http://www.arXiv.org/abs/hep-ph/0604232}{{\tt hep-ph/0604232}}.

\bibitem{sherpa}
T.~Gleisberg {\em et al.}, {\em J. High Energy Phys.} {\bf 02} (2009) 007,
  \href{http://www.arXiv.org/abs/hep-ph/0811.4622}{{\tt hep-ph/0811.4622}}.

\bibitem{sherpa-2.2}
E.~Bothmann {\em et al.}, {\em SciPost Phys.} {\bf 7} (2019) 034,
  \href{http://www.arXiv.org/abs/hep-ph/1905.09127}{{\tt hep-ph/1905.09127}}.

\bibitem{jad-yost-afb}
S.~Jadach and S.~Yost, {\em Phys. Rev. D} {\bf 100} (2019) 013002,
  \href{http://www.arXiv.org/abs/1801.08611}{{\tt 1801.08611}}.

\bibitem{ichep2022-say}
{S.A. Yost} {\em et al.}, {\em PoS} {\bf ICHEP2022} (2022) 887,
  \href{http://www.arXiv.org/abs/2211.17177}{{\tt 2211.17177}}.

\bibitem{sjetaltoappear}
S.~Jadach {\em et al.}, {to appear}.

\bibitem{kkmchh2}
S.~Jadach, B.~F.~L. Ward, Z.~W\c{a}s, and S.~Yost,
  \href{http://www.arXiv.org/abs/hep-ph/2002.11692}{{\tt hep-ph/2002.11692}}.

\bibitem{Jadach:2018jjo}
S.~Jadach, W.~P\l{}aczek, M.~Skrzypek, B.~F.~L. Ward, and S.~A. Yost, {\em
  Phys. Lett. B} {\bf 790} (2019) 314--321,
  \href{http://www.arXiv.org/abs/1812.01004}{{\tt 1812.01004}}.

\bibitem{Jadach:2021ayv-sh}
S.~Jadach, W.~P\l{}aczek, M.~Skrzypek, and B.~F.~L. Ward, {\em Eur. Phys. J. C}
  {\bf 81} (2021) 1047.

\bibitem{fcc2023wkshpms}
M.~Skrzypek {\em et al.}, talk in {\it 2023 FCC Workshop, Krakow, PL}.

\bibitem{fjeger-fccwksp2019}
F.~Jegerlehner, {\em {CERN Yellow Reports: Monographs, eds. A. Blondel {\it et
  al.}, CERN-2020-003}} (2020) 9,
  \href{http://www.arXiv.org/abs/1905.05078}{{\tt 1905.05078}}.

\bibitem{latt1}
S.~Borsanyi {\em et al.},
  \href{http://www.arXiv.org/abs/hep-lat/1711.04980}{{\tt hep-lat/1711.04980}}.

\bibitem{latt2}
M.~Ce {\em et al.}, \href{http://www.arXiv.org/abs/hep-lat/2203.08676}{{\tt
  hep-lat/2203.08676}}.

\bibitem{ward:2013dkunv}
B.~F.~L. Ward, {\em Phys. Dark Universe} {\bf 2} (2013) 97 -- 109.

\bibitem{ijmpa2018}
{B.F.L. Ward}, {\em Int. J. Mod. Phys. A} {\bf 33} (2018) 1830028.

\bibitem{reuter2}
{A. Bonanno and M. Reuter}, {\em Jour. Phys. Conf. Ser.} {\bf 140} (2008)
  012008.

\bibitem{pdg2008}
C.~Amsler {\em et al.}, {\em Phys. Lett. B} {\bf 667} (2008) 1.

\bibitem{sola2}
J.~Sola, {\em J. Phys. A} {\bf 41} (2008) 164066.

\bibitem{sola3}
J.~Sola~Peracaula, \href{http://www.arXiv.org/abs/hep-ph/2308.13349, and
  references therein}{{\tt hep-ph/2308.13349, and references therein}}.

\bibitem{Jadach:2023cl1}
S.~Jadach, B.~F.~L. Ward, and Z.~W\c{a}s,
\href{http://www.arXiv.org/abs/hep-ph/2303.14260}{{\tt hep-ph/2303.14260}}.

\bibitem{gribv-lptv:1972}
V.~Gribov and L.~Lipatov, {\em Sov. J. Nucl. Phys.} {\bf 15} (1972), no.~9 675,
  938.

\bibitem{berends-neerver-burgers:1988}
F.~Berends, W.~V. Neerven, and G.~Burgers, {\em Nucl. Phys.} {\bf B297} (1988)
  429; Erratum: Nucl. Phys. {\bf B304} (1988) 921 -- 922.

\bibitem{ceex2:1999sh}
S.~Jadach, B.~F.~L. Ward, and Z.~W\c{a}s, {\em Phys. Rev. D} {\bf 63} (2001)
  113009.

\bibitem{gps:1998}
S.~Jadach, Z.~W\c{a}s, and B.~F.~L. Ward, {\em Eur. Phys. J.} {\bf C22} (2001)
  423--430, \href{http://www.arXiv.org/abs/hep-ph/9905452}{{\tt
  hep-ph/9905452}}.

\bibitem{bflwetaltoappear}
B.~F.~L. Ward {\em et al.}, {to appear}.

\end{thebibliography}\endgroup
\bibliographystyle{utphys_spires}

\end{document}